\documentclass[11pt]{article}
\usepackage{moriond,epsfig,amsmath,amssymb}

\def\Journal#1#2#3#4{{#1} {\bf #2}, #3 (#4)}

\def\NPB{{\em Nucl. Phys.} B}
\def\PLB{{\em Phys. Lett.}  B}
\def\PRL{\em Phys. Rev. Lett.}
\def\PRD{{\em Phys. Rev.} D}

\def\be{\begin{equation}}
\def\ee{\end{equation}}
\def\bea{\begin{eqnarray}}
\def\eea{\end{eqnarray}}

\begin{document}
\vspace*{4cm}
\title{$\Delta F = 1$ CONSTRAINTS ON MINIMAL FLAVOR VIOLATION}

\author{J. F. KAMENIK}

\address{INFN Laboratori Nazionali di Frascati,\\
Via E. Fermi 40, I-00044 Frascati, Italy,\\
J. Stefan Institute,\\
Jamova 39, P. O. Box 3000, 1001 Ljubljana, Slovenia}

\maketitle\abstracts{
We present an updated phenomenological analysis of the minimal flavor violating (MFV) effective theory, both at small and large $\tan \beta$, in the sector of $\Delta F=1$ processes. We evaluate the bounds on the scale of new physics derived from recent measurements (in particular from $B \to X_s \gamma$, $B \to X_s \ell^+ \ell^-$, $B_s \to \mu^+ \mu^-$ and $K\to\pi\nu\bar\nu$) and we use such  bounds to derive a series of model-independent predictions within MFV for future experimental searches in the flavor sector.
}

\section{Introduction}

The Standard Model (SM) accurately describes high energy physical phenomena up to the electro-weak (EW) scale $\mu_W \sim 100$~GeV. It is however known to be incomplete due to the lack of description of gravity, proper unification of forces as well as neutrino masses. In view of these shortcomings, it can be regarded as a low-energy effective description of physics below a UV cut-off scale $\Lambda$. But if it is an effective theory, at what scale $\Lambda$ below the unification or the Planck scale does it break down? 
%Using a bottom-up approach, the answer can be sought by examining the dimensionful parameters of the effective SM Lagrangian, expanded at the $\mu_W$ scale in a series of local operators of increasing dimension. 
The only dimensionful parameter in the renormalizable part of the Lagrangian is the Higgs mass, which is known to be quadratically sensitive to the cut-off scale of the theory. Then the EW hierarchy problem suggests that new physics (NP) should appear around or below $\Lambda\lesssim 1$~TeV.  The non-renormalizable higher dimensional terms, formally suppressed by the increasing powers of the cut-off scale on the other hand mediate flavor changing neutral currents (FCNCs), may contain additional sources of CP violation and can violate baryon and lepton numbers. Even in absence of the later, precision measurements of low energy experiments put severe constraints on the scale of flavor and CP violating NP. Excellent agreement between SM predictions and experiment on $\epsilon_K$ (constraining $s-d$ sector), $A_{CP}(B_d\to \Psi K_s)$ and $\Delta m_d$ (in the $b-d$ sector) and $B\to X_s\gamma$ (for $b\to s$ transitions) constrains a general flavor violating NP to appear above $\Lambda\gtrsim 2\times10^5$~TeV, $~2\times 10^3$~TeV and $40$~TeV respectively. The resulting tension between the two estimates of the NP scale illustrates what is often called the new physics flavor problem.
%The dots stand for terms supressed by higher powers of the $\Lambda$ scale. 

\par

The Minimal Flavor Violation (MFV) hypothesis~\cite{ref:mfv-others,ref:mfv-gino} aims to solve the issue by demanding that all flavor symmetry breaking in and also beyond the SM is proportional to the SM Yukawas. A few direct consequences follow from this assumption: Firstly the Cabibbo-Kobayashi-Maskawa (CKM) matrix is the only source of flavor mixing and CP violation even beyond the SM. Thus, all (non-helicity suppressed) tree level and CP violating processes are constrained to their SM values. Finally, CKM unitarity is maintained and a (universal) unitarity triangle (UUT) can be determined from a constrained set of observables~\cite{ref:uut}. Other details of phenomenology depend on the form of the EW Higgs sector of the theory. In case of a SM-like single Higgs doublet, the FCNCs in the down quark sector are all driven by the large top Yukawa ($\lambda_t$). At the same time, when performing the operator product expansion (OPE) at the EW scale, the SM basis of operators contributing to the effective weak Hamiltonian is complete also in presence of NP, making the MFV effective theory approach predictive. The same holds true at low $\tan\beta\equiv v_u/v_d$ if the Higgs sector is described in terms of an effective two Higgs doublet model of type II with the vacuum expectation values of the Higgses coupling to up(down) quarks denoted by $v_{u(d)}$. However, bottom Yukawa ($\lambda_b$) contributions become important at large $\tan\beta$ as $\lambda_b (\sim m_b \tan \beta/v_u) \sim \lambda_t$. Accompanied by the partial lifting of helicity suppression in the down sector, contributions due to new density operators have to be taken into account in the effective weak Hamiltonian. Still, the predictivity of the MFV effective theory approach is maintained by the small number of additional operators which need to be considered.

\par

The symmetry principles underlying the MFV hypothesis establish solid links among different flavor observables at low energy and allow to probe and constrain the scale of MFV NP. Since (non-helicity suppressed) charged current interactions are not affected, bounds can be derived from $\Delta F=2$ and $\Delta F=1$ FCNC phenomenology. The $\Delta F=2$ processes are box loop mediated in the SM, and only a few operators contribute to the effective weak Hamiltonian. The main observables here are the $K$, $B_q$ oscillation parameters to which MFV NP at low $\tan\beta$ contributes universally~\cite{ref:mfv-gino}. A recent analysis~\cite{ref:utfit} was able to constrain this contribution and put a lower bound on the effective NP scale $\Lambda>5.5$~TeV at $95\%$ probability. The $\lambda_b\tan\beta$ contributions break the universality among kaon and $B$ meson sectors at large $\tan\beta$, resulting in a slightly weaker bounds of $\Lambda>5.1$~TeV. New operators due to Higgs exchange in the loop start contributing only at very large values of $\tan\beta$, resulting in a bound on a certain combination of charged Higgs parameters. $\Delta F=1$ processes on the other hand are penguin loop mediated in the SM, with many operators contributing. In concrete MFV models, they are often related to the $\Delta F=2$ as well as flavour conserving phenomenology~\cite{ref:haisch}. On the other hand in our effective theory bottom-up approach they have to be considered completely orthogonal. An analysis of bounds coming from radiative, and (semi)leptonic decays of $K$ and $B$ mesons was performed a while ago~\cite{ref:mfv-gino}, however limited experimental information at the time barred from exploring in particular the interesting role of the large $\tan\beta$ scenario. In the meantime, the situation has drastically improved and the new updated experimental and theoretical results on $\Delta F=1$ FCNC mediated processes further motivate the revisiting and updating of this analysis. In the following we present a selection of results from such a study, the details of which will be presented elsewhere~\cite{ref:mfv-new}.

\section{Updating Analysis of $\Delta F=1$ Constraints}

In the SM the effective weak Hamiltonian describing $\Delta F=1$ FCNC processes among down-type quark flavors $q_i-q_j$ can be written as~\cite{ref:mfv-gino}
\begin{equation}
\mathcal{H}^{\Delta F=1}_{eff} = {\frac{G_{F} \alpha_{\mathrm{em}}}{
2\sqrt{2}\pi \sin^2 \theta_{W}}
 V^*_{ti} V_{tj}} \sum_{n}{C_n}
{\cal Q}_n~+ ~{\rm h.c.}\,,
\end{equation}
where $G_F$ is the Fermi constant, $\alpha_{\mathrm{em}}$ is the fine structure constant, $\theta_W$ is the Weinberg angle and $V_{ij}$ are the CKM matrix elements. The short distance SM contributions are encoded in the Wilson coefficients $C_n$, computed via perturbative matching procedure at the EW scale. MFV NP manifests itself in the shifts of the individual Wilson coefficients in respect to the SM values $C_n(\mu_W) = C_n^{SM} + \delta C_n$. These shifts can be translated in terms of the tested NP energy scale $\Lambda$ as $\delta C_n = 2 a \Lambda_0^2/\Lambda^2$, where $\Lambda_0=\lambda_t \sin^2(\theta_W) m_W/\alpha_{em} \sim 2.4$~
TeV is the corresponding typical SM effective energy scale. The value of the free variable $a$ depends on the details of a particular MFV NP model. In general $a\sim 1$ for tree level NP contributions, while $a\sim 1/16\pi^2$ for loop suppressed NP contributions. In our numerical results we put $a$ to unity. 

In order to address low energy phenomenology, one needs to evaluate the appropriate matrix elements of the corresponding effective dimension 6 operators $\mathcal Q_n$. At low $\tan\beta$ we consider the electro-magnetic (EM) and QCD dipole operators
\begin{equation}
\begin{array}{rclrcl}
%\multicolumn{6}{c}{\mathrm{EM~and~QCD~dipole~operators}}\\
{\cal Q}_{7\gamma} &=& \frac{2}{g^2}  m_{j} \bar d_{iL} \sigma_{\mu\nu} d_{jR} (e F_{\mu\nu})\,, & {\cal Q}_{8G} &=&\frac{2}{g^2}  m_{j} \bar d_{iL} \sigma_{\mu\nu} T^a  d_{jR}  (g_s G^a_{\mu\nu})\,,\\
\end{array}
\end{equation}
where $g$ is the EW $SU(2)_L$ coupling, $e$ is the EM coupling, $g_s$ is the QCD coupling, $T^a$ are the $SU(3)_c$ generator matrices, while $F_{\mu\nu}$ and $G_{\mu\nu}^a$ are the EM and QCD field tensors. They contribute to $B\to X_s\gamma$ decay as well as to the $B\to X_s\ell^+\ell^-$ phenomenology, where in addition we get contributions from the EW-penguin operators
\begin{equation}
\begin{array}{rclrcl}
%\multicolumn{6}{c}{{\mathrm{EW-penguin~operators}}}\\
{\cal Q}_{9V}   &=& 2 \bar d_{iL} \gamma_\mu d_{jL}~ \bar \ell \gamma_\mu \ell\,, & {\cal Q}_{10A}  &=& 2 \bar d_{iL} \gamma_\mu d_{jL} ~  \bar \ell \gamma_\mu \gamma_5 \ell\,.
\end{array}
\end{equation}
Here $\ell = e,\mu,\tau$ denotes the charged leptons. $\mathcal Q_{10A}$ also mediates $B_q\to \ell^+\ell^-$. Finally the Z-penguin operator
\begin{equation}
{\cal Q}_{\nu\bar\nu} = 4 \bar d_{iL} \gamma_{\mu} d_{jL} \bar \nu_L \gamma_{\mu} \nu_L
\end{equation}
enters solely in $B\to X_s\nu\bar\nu$ and $K\to\pi\nu\bar\nu$ decays and can thus be constrained independently of the others. We do not consider NP contributions to QCD penguin operators as their impact on phenomenology is subdominant compared to long distance effects. At large $\tan\beta$, one needs to take into account an additional density operator 
\begin{equation}
{{\cal Q}_{S-P}} = {4 (\bar d_{iL} d_{jR}) (\bar \ell_R \ell_L)}
\end{equation}
contributing to $B\to X_s\ell^+\ell^-$ and $B_q\to \ell^+\ell^-$. On the other hand, contributions from additional four quark density operators~\cite{ref:HillerKruger}~\footnote{We thank Ulrich Haisch for pointing out these potential contributions.} which are also $\tan\beta$ enhanced and enter $B\to X_s\gamma$ and $B\to X_s\ell^+\ell^-$ through one loop mixing with $\mathcal Q_{7\gamma,8G}$ are $\alpha_{\mathrm{em}}/4\pi\sim 0.001$ suppressed relative to those of ${{\cal Q}_{S-P}}$ and thus turn out to be negligible after imposing the bounds on ${{\cal Q}_{S-P}}$.

In our analysis we consider the most theoretically clean observables in order to derive reliable bounds on possible NP contributions. In particular, we use the inclusive branching ratio of the radiative $B\to X_s\gamma$ decay, measured with a lower cut on the photon energy. The latest HFAG value averaged over different measurements~\cite{ref:bsg-hfag}
%~\footnote{Not yet including the latest update from Belle~\cite{ref:Belle-bsg}} 
is $Br(B\to X_s\gamma)_{E_{\gamma}>1.6~\mathrm{GeV}}^{\mathrm{exp}}=3.52(23)(9) \times 10^{-4}$, where the first error is statistical and the second systematic. Theoretically, the SM value is known to better than $8\%$ and the expansion in terms of $\delta C_n$ evaluated at the weak scale is~\cite{ref:Misiak-bsg}
\begin{eqnarray}
\label{eq:bsg}
Br(B\to X_s\gamma)_{E_{\gamma}>1.6~\mathrm{GeV}}^{\mathrm{th}}&=&3.16(23)\left(1 - {2.28}   {\delta C}_{{7\gamma}} - {0.71}   {\delta C}_{{8G}} \right.\nonumber\\
&&\left.\hspace{1.4cm} + {1.51} {\delta C}_{{7\gamma}}^2+{0.78} {\delta   C}_{{8G}} {\delta   C}_{{7\gamma}}+{0.25} {\delta   C}_{{8G}}^2\right) \times 10^{-4}\,,
\end{eqnarray}
where the central value and its error have been adjusted to take into account the CKM matrix element determination from the UUT analysis~\cite{ref:utfit}. Since $\delta C_7$ and $\delta C_8$ in absence of four quark density operator contributions enter in the same fixed combination to all relevant observables (any differences being artifacts of the truncated perturbative expansion) one can always eliminate one of them (e.g. $\delta C_{8G}$) from the analysis and then reconstruct the bound on both from the quadratic combination in eq~(\ref{eq:bsg}). 

A completely different combination of operators contributes to the helicity suppressed decay $B_s\to\mu^+\mu^-$. Experimentally the best upper bound on the branching ratio was recently put by the CDF collaboration~\cite{ref:CDF-Bsmumu} $Br(B_s\to\mu^+\mu^-)^{\mathrm{exp}}<4.7\times10^{-8}$ at $90\%$ C.L.\,, which is only an order of magnitude above the SM prediction. The theoretical error of which is around $23\%$ and is dominated by the lattice QCD determination of the $B_s$ decay constant.%~\footnote{When considering specific MFV models (also the SM), the theoretical error can be reduced by combining the data on $\Delta m_s$~\cite{ref:Buras-Bsmumu}. Such correlations are lost in the effective theory approach adopted here.} 
Again using UUT CKM inputs, the expansion in terms of $\delta C_i$ reads
\begin{eqnarray}
Br(B_s\to\mu^+\mu^-)^{\mathrm{th}}&=&3.8(9)\left( 1-{2.1} {\delta C}_{{10A}}-{2.3}   {\delta C}_{{S-P}}\right.\nonumber\\
&&\hspace{1.2cm}\left.+ {1.1} {\delta C}_{{10A}}^2+{2.4} {\delta C}_{{S-P}} {\delta C}_{{10A}}+{2.7}   {\delta C}_{{S-P}}^2 \right)\times 10^{-9}\,.
\end{eqnarray}

Analysis of $B\to X_s\ell^+\ell^-$ is more involved since, not only do almost all of the above mentioned operators ($\mathcal Q_{7\gamma,8G,9V,10A,S-P}$) contribute here, experimentally there are already a number of inclusive as well as exclusive measurements available, constraining different combinations of NP parameters. On the inclusive side, only the branching ratio $Br(B\to X_s\ell^+\ell^-)$, where $\ell=e,\mu$ is measured by the B factories~\cite{ref:BtoXsll-exp} in several bins of di-lepton invariant mass squared ($q^2$). The errors vary from almost $90\%$ in the first bin where only Belle has obtained a relevant signal, to around $30\%$ in the other bins. 
%Theoretically, the charmonium region has to be excluded due to non-perturbative effects,
%while in the highest $q^2$ region, EM radiative corrections become important. 
%while non-perturbative effects in the first bin reduce the precision to around $30\%$~\cite{ref:BtoXsll-?}. Unfortunately this region is most sensitive to contributions of $C_{7\gamma}$, which scale as $1/q^2$ relative to other contributions. 
The latest calculations estimate the theoretical error at around $7\%$ for the bins below the charmonium region and around $10\%$ for the high $q^2$ bin~\cite{ref:BtoXsll-Hurth}. The relevant formulae including NP contributions are rather lenghty and can be found in ref.~\cite{ref:mfv-new,ref:BtoXsll-Hurth}. 

Much more experimental information is available for exclusive channels where the $B\to K^{(*)}\ell^+\ell^-$ branching ratios as well as several angular distributions have already been measured~\cite{ref:BtoKll-exp}. Theoretically however, despite considerable theoretical progress on the evaluation of the non-perturbative matrix elements of $\mathcal Q_n$ entering exclusive channels in the recent years~\cite{ref:scet}, a reliable determination can only be expected from fundamentally non-perturbative methods, such as lattice QCD. In the meantime, any phenomenological implications based on existing form factor estimates should be treated with care. We will present an analysis of the impact of the exclusive modes on the MFV NP bounds elsewhere~\cite{ref:mfv-new}.

%These are important as they probe different combinations of Wilson coefficients than the decay rates. Theoretically, these observables can be computed in terms of the non-perturbative matrix elements of $\mathcal Q_n$, whose structure is then encoded by sets of scalar functions of $q^2$ -- the form factors. Although there was considerable theoretical progress on the evaluation of the form factors using QCD sum rules (QCDSR)~\cite{ref:qcdsr} and soft collinear effective theory~\cite{ref:scet}, a reliable determination can only be expected from fundamentally non-perturbative methods, such as lattice QCD. In the meantime, any phenomenological implications based on existing form factor estimates should be treated with care -- the associated theoretical errors should be assigned conservatively. We will present such an analysis of the impact of the exclusive modes on the MFV NP bounds elsewhere~\cite{ref:mfv-new}.

Finally MFV NP contributions to the Z-penguin operators can be constrained using the first experimental hints~\cite{ref:kpinunu-exp} of the $K^+\to \pi^+\nu\bar\nu$ decay $Br(K^+\to \pi^+\nu\bar\nu(\gamma))^{\mathrm{exp}}=147(120)\times 10^{-12}$ and comparing them to the theoretical predictions, which are brought under control by the use of experimental data on $K\ell3$ decays~\cite{ref:mescia-smith} resulting in only $11\%$ theoretical error. In presence of MFV NP the corresponding expression reads
\begin{equation}
Br(K^+\to \pi^+\nu\bar\nu(\gamma))^{\mathrm{th}}=7.53(82)(1+0.93\delta C_{\nu\bar\nu} + 0.22\delta C_{\nu\bar\nu}^2)\times 10^{-11}\,.
\end{equation}

Common parametric inputs in our analysis are the particle masses and lifetimes from PDG~\cite{ref:pdg} as well as the parameters of the CKM matrix, which, as already mentioned, we take from the UUT analysis~\cite{ref:utfit}. We perform a correlated fit of subsets of observables turning on NP contributions and extract probability bounds on the shifts of the Wilson coefficients away from their SM values. 
%The details of the procedure will be presented elsewhere~\cite{ref:mfv-new}. 

\section{Results}

\begin{figure}[t]
\begin{center}
\hspace{-0.7cm}
\epsfxsize5.3cm\epsffile{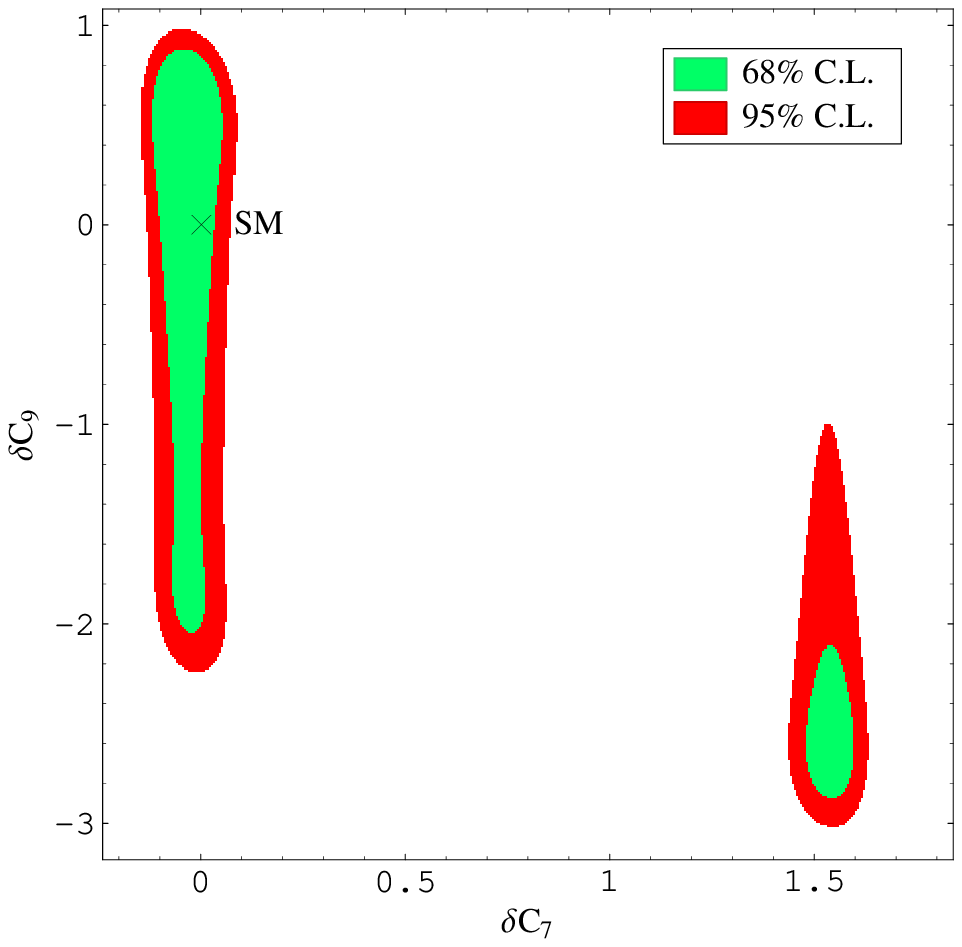}
\epsfxsize5.3cm\epsffile{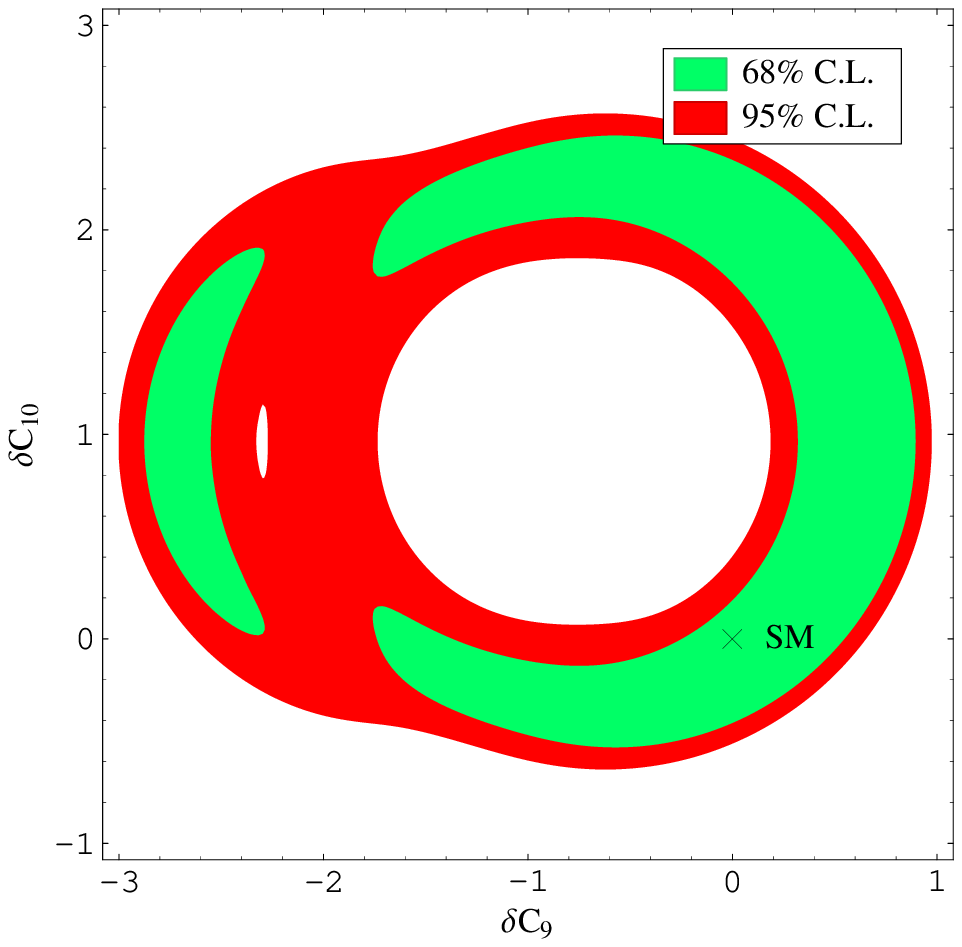}
\epsfxsize5.3cm\epsffile{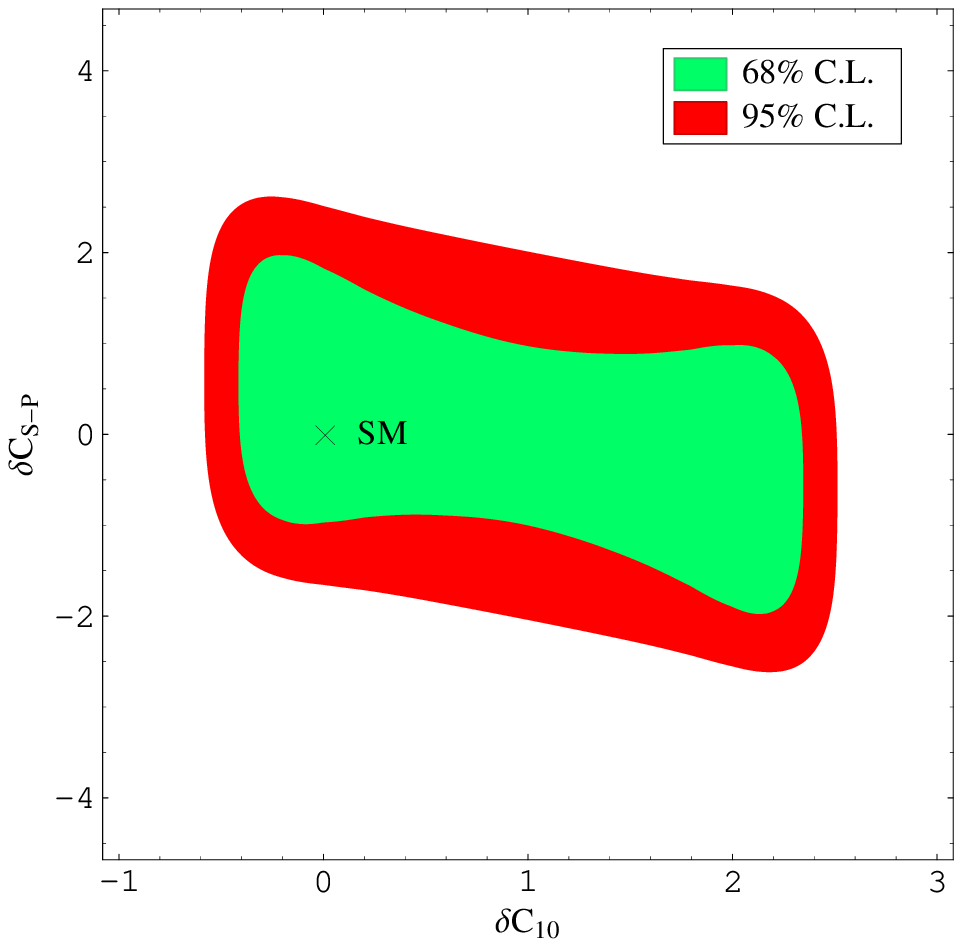}
\end{center}
\caption{\label{fig:corelations} Correlation plots showing the most pronounced correlations among the bounds on the various NP Wilson coefficient shifts. The $68\%$ ($95\%$) probability regions are shown in green (red).}
\end{figure}

The compilation of bounds on the MFV NP scale in respect to all the probed operators is summarized in table~\ref{table:lambda-bounds}.
\begin{table}[t]
\caption{\label{table:lambda-bounds}Summary of bounds on the MFV NP scales related to the probed effective operators. All the numerical values are the lower bounds at $95\%$ probability on the MFV NP scale $\Lambda$ as explained in the text.}
\vspace{0.4cm}
\begin{center}
\begin{tabular}{|c|c|c|}
\hline
Operator & Conservative bound~[TeV] & Natural bound~[TeV]  \\ \hline
$\mathcal Q_{7\gamma}$ & $1.6$ & $5.3$ \\
$\mathcal Q_{8G}$ & $1.2$ & $3.1$ \\
$\mathcal Q_{9V}$ & $1.4$ & $1.6$ \\
$\mathcal Q_{10A}$ & $1.5$ & $1.5$ \\
$\mathcal Q_{S-P}$ & $1.2$ & / \\
$\mathcal Q_{\nu\bar\nu}$ & $1.5$ & / \\ \hline
\end{tabular}
\end{center}
\end{table}
We present two sets of bounds. In the conservative estimate we take into account all the possible fine-tunings and cancellations among the various operator contributions, including discrete ambiguities in cases where the NP contributions might flip the sign of the SM pieces. For the second, more natural bounds, we consider each $\delta C_n$ individually and also discard flipped-sign fine-tunned solutions. The strongest bounds come naturally from the $B\to X_s\gamma$ decay rate and affect $\mathcal Q_{7\gamma,8G}$. As can be seen, the effect of the discrete ambiguity is large and only the natural bounds on $\Lambda>5.2(3.1)$ for $Q_{7\gamma(8G)}$ are competitive with the ones on $\Delta F=2$ operators~\cite{ref:utfit}. 
The discrete ambiguity (also seen on utmost left plot in figure~\ref{fig:corelations}) could however be completely removed in the future once the experimental situation concerning the lowest $q^2$ region in $B\to X_s\ell^+\ell^-$ rate and especially the forward-backward asymmetry (FBA) improves. 
As expected, $\mathcal Q_{S-P,\nu\bar\nu}$ operators are mainly bounded from single observables ($B_s\to\mu^+\mu^-$ and $K^+\to\pi^+\nu\bar\nu$ respectively) leading to robust bounds around $1.2$~TeV and $1.5$~TeV respectively. Finally $\delta C_{9V,10A}$ are mainly bounded by $B\to X_s\ell^+\ell^-$ and using only presently available inclusive information the bounds are around $1.5$~TeV. In all of the considered observables except $B\to X_s\gamma$ the experimental uncertainties strongly dominate and at present do not allow to discern discrete ambiguities or strong correlations as can be also deduced from figure~\ref{fig:corelations} showing the most interesting pairwise correlation plots of the $68\%$ and $95\%$ allowed parameter regions.

\section{Discussion and Outlook}

In summary, immense experimental and theoretical progress in the area of flavor physics in the last decade has made it possible to constrain in a model independent way the complete set of possible beyond SM contributions to $\Delta F=1$ and $\Delta F=2$ processes due to possible MFV NP both at small and large $\tan\beta$. Bounds coming from $\Delta F=2$ phenomenology are already very constraining, pushing the effective MFV NP scale beyond $5$~TeV. In $\Delta F=1$ sector, at present only the bounds coming from $B\to X_s\gamma$ are of comparable strength. However most uncertainties are dominated by experiments and one can look forward for the results of full dataset analyses by the B factories. 

Using the derived bounds on the MFV NP contributions in $\Delta F=1$ processes we are able to make predictions for other potentially interesting observables to be probed at LHCb or a future Super Flavor Factory. As already mentioned, angular distributions like the FBA probe different combinations of the operators and would provide complimentary bounds. At the moment, considering bounds from inclusive measurements alone, no firm constraints on the FBA or its zero can be be imposed within MFV models. This conclusion reinforces the importance of these observables and their potentiality of discovering relevant deviations. 
%Existing information coming from exclusive modes will be discussed elsewhere~\cite{ref:mfv-new}.

Another set of observables displays interesting sensitivity to the $\tan\beta$ enhanced $C_{S-P}$ contributions. Such are lepton flavor universality ratios $\Gamma(B\to K^{(*)}\mu^+\mu^-)/\Gamma(B\to K^{(*)}e^+e^-)$, which are very close to 1 with the SM as well as MFV models with low $\tan\beta$. However even at $\large\tan\beta$ present constraints already disallow deviations larger then $10\%$ from unity for such ratios.

Finally the derived bounds allow to construct tests able to potentially rule out MFV. Beside the interesting CP violation signals already emerging in the $B_s$ sector~\cite{ref:cpv-bs}, in $\Delta F=1$ sector first there are the firm relations among the different flavor transitions [$(b \leftrightarrow s)/(b \leftrightarrow d)/(s \leftrightarrow d) \sim |V_{tb}V_{ts}^*|/ |V_{tb}V_{td}^*| / |V_{ts}V_{td}^*|$] which might be probed with $K\to \pi \ell^+\ell^-$, $B\to X_s \nu\bar\nu$ or $B_d\to\mu^+\mu^-$ processes. Also interesting in this respect is the FBA in $B\to K\ell^+\ell^-$ which is already restricted to be below $1\%$ within MFV models regardless of $\tan\beta$.

\section*{Acknowledgments}
{\footnotesize
The author wishes to thank the organizers of the 
Rencontres de Moriond EW 2008 for their invitation and hospitality, as well as the collaborators on this project T. Hurth, G. Isidori and F. Mescia for discussions and comments on the manuscript. An Experienced Researcher position supported by the EU-RTN Programme,
Contract No. MRTN--CT-2006-035482, \lq\lq Flavianet'' is gratefully acknowledged. 
}

\end{document}